\documentclass[amstex,useAMS,epsf,amssymb,usenatbib]{mn2e}
\usepackage{epsfig}
\usepackage{graphicx}
\usepackage{color}
\usepackage{amssymb,amsmath}
\usepackage{comment}
\usepackage{hyperref}

\usepackage{soul}
\usepackage{xcolor}
\setulcolor{cyan} 
\setstcolor{red}     
\sethlcolor{yellow} 

\def\simlt{\ \raise -2.truept\hbox{\rlap{\hbox{$\sim$}}\raise5.truept   %
\hbox{$<$}\ }}
\def\simgt{\ \raise -2.truept\hbox{\rlap{\hbox{$\sim$}}\raise5.truept   %
\hbox{$>$}\ }}                                                          %

\title[The first measurement of temperature standard deviation]{The first measurement of temperature standard deviation
along the line-of-sight in galaxy clusters}

\author[D. A. Prokhorov and S. Colafrancesco]{D. A. Prokhorov$^{1}$\thanks{E-mail:
phdmitry@stanford.edu}, S. Colafrancesco$^{2,3}$
\\
~\\ $^{1}$W. W. Hansen Experimental Physics Laboratory, Stanford
University, Stanford, CA 94305, USA\\ $^{2}$ University of the
Witwatersrand, School of Physics, Private Bag 3 - Wits 2050,
Johannesburg, South Africa\\
 $^{3}$ INAF - Osservatorio Astronomico di Roma, via Frascati 33, I-00040 Monteporzio, Italy.}

\date{Accepted .....
      Received ..... ;
      in original form .....}

\pagerange{\pageref{firstpage}--\pageref{lastpage}}
\pubyear{2011}

\begin{document}

\maketitle

\label{firstpage}

\begin{abstract}

Clusters of galaxies are mainly formed by merging of smaller
structures, according to the standard cosmological scenario. If
the mass of a substructure is $\gtrsim 10\%$ of that of a galaxy
cluster, the temperature distribution of the intracluster medium
(ICM) in a merging cluster becomes inhomogeneous. Various methods
have been used to derive the two-dimensional projected temperature
distribution of the ICM. However, methods for studying temperature
distribution along the line-of-sight through the cluster were
absent. In this paper, we present the first measurement of the
temperature standard deviation along the line-of-sight, using as a
reference case the multifrequency SZ measurements of the Bullet
Cluster. We find that the value of the temperature standard
deviation is high and equals to (10.6$\pm$ 3.8) keV in the Bullet
Cluster. This result shows that the temperature distribution in
the Bullet Cluster is strongly inhomogeneous along the
line-of-sight and provides a new method for studying galaxy
clusters in depth.

\end{abstract}


\section{Introduction}

Clusters of galaxies are the largest gravitationally bound
structures with size $\sim$1 $\div$ 3 Mpc ($\approx$ (3 $\div$
10)$\times10^{24}$ cm) containing a hot diffuse plasma (the
intracluster medium or ICM) which eventually sets in
equilibrium in the potential wells of the cluster. The plasma is
hence heated to high temperatures during the cluster gravitational
collapse and the subsequent virialization by the gravitational
energy released by the cluster formation from the assembly of
smaller structures. The temperature of the ICM in its equilibrium
stage can be as high as $10 \div 15$ keV in the most massive
clusters and ensures that the elements present in the ICM are
highly ionized. Abundant light elements (such as hydrogen, helium,
and oxygen) in the ICM have all the electrons removed from their
nuclei. The free hot electrons in the ICM cause X-ray
bremsstrahlung emission due to the interaction with protons (see,
e.g., Rybicki \& Lightman 1979) and also cause a distortion of the
CMB radiation towards galaxy clusters due to the Inverse Compton
(IC) scattering off CMB photons (the Sunyaev-Zel'dovich effect,
hereafter the SZ effect; see, e.g., Sunyaev-Zel'dovich 1980). The
typical number density of free electrons in the ICM is $10^{-3}
\div 10^{-2}$ cm$^{-3}$.

Mergers of sub-clusters occurring during cluster evolution cause
an additional significant and non-uniform heating of the ICM due
to the conversion of kinetic energy of gravitationally accelerated
sub-clusters into thermal energy released into the ICM by shock
waves. Shocks have been found, in fact, by the Chandra X-ray
observatory in various systems, from galaxy groups with
temperature of $\approx$1 keV to the most massive galaxy clusters
with temperature of $\approx$15 keV (for a review, see Markevitch
\& Vikhlinin 2007). The cluster 1ES0657-558 (known as the Bullet
Cluster) is a ``textbook''  example of a hot, merging cluster
which has a bullet-like substructure in the ambient ICM, a strong
shock wave driven by the bullet substructure, and very high and
non uniform values of temperature (Markevitch et al. 2002). Thanks
to deep Chandra observations, the Bullet Cluster is one of the
best-studied galaxy clusters in X-rays. The analysis of the X-ray
data from Chandra (see, e.g.,Million \& Allen 2008) shows that a
single temperature model (i.e., a single temperature value
observed along the line-of-sight through the cluster) does not
properly describe the observed its X-ray spectrum. The temperature
distribution along the line-of-sight in this cluster is then
presumably strongly inhomogeneous.

Gas temperature in galaxy clusters can be derived by using
different  techniques, such as analysing of the X-ray spectra or
the spectral form of the CMB distortion owing to the SZ effect.
The derived X-ray spectroscopic temperature for a multitemperature
electron population is weighted with a factor that is proportional
to the squared number density and to the gas temperature to the
power $\approx$-3/4 (Mazzotta et al. 2004). Therefore, due to the
non-linear combination of density and temperature there is no
simple way to interpret the X-ray spectroscopic temperature as the
actual gas temperature and to obtain direct information about the
temperature distribution along the line-of-sight. Additional
assumptions about the gas density and temperature profiles along
the line-of-sight, which are \textit{a priori} unknown in merging
galaxy clusters, are required for the X-ray analysis (see, e.g.,
Samsing et al. 2012). The temperature derived through the SZ
technique is the mass-weighted temperature (Hansen 2004; Prokhorov
et al. 2010, see also Colafrancesco \& Marchegiani 2010 for a
discussion on the spectroscopic SZ temperature derivation) and,
therefore, it has a more explicit interpretation. Below, we will
use this fact to obtain the information about the temperature
distribution along the line-of-sight.

The SZ effect is an important tool for studying of galaxy
clusters, since both the amplitude and spectral form of the CMB
distortion (caused by the IC scattering of CMB photons off free
electrons) depend on the ICM physical parameters (e.g., the plasma
pressure and temperature). Relativistic corrections to the SZ
effect are significant for high-temperature plasmas in massive
clusters of galaxies (see Rephaeli 1995). The contribution from
the relativistic corrections becomes stronger at high frequencies.
The first detection of the SZ effect brightness increment from the
Bullet Cluster at frequencies above 600 GHz has been obtained with
the Herschel-SPIRE instrument (Zemcov et al. 2010). These
high-frequency SZ observations have been used by Colafrancesco et
al. (2011) to demonstrate that plasma models with several electron
components provide a better fit than that with a single thermal
component. This means that a substantial temperature variance (or
standard deviation) $\sigma$ is expected along the line of sight
through this cluster.
The method of measurements of the temperature standard deviation,
$\sigma$, along the line-of-sight has been proposed by Prokhorov
et al. (2011). The use of the temperature standard deviation,
which is a measure of the plasma inhomogeneity along the line of
sight, provides us hence with a model-independent
approach to characterize the amount of the temperature
inhomogeneity in the ICM.

In this paper, we present the first measurement of the temperature
standard deviation $\sigma$ along the line-of-sight in galaxy
clusters. We use the method of Prokhorov et al. (2011) and SZ
effect intensity data for the Bullet Cluster measured at four
frequencies (150, 275, 600, and 857 GHz) to derive the temperature
standard deviation. Using this method, we demonstrate that the
temperature distribution along the line of sight through this
cluster is very inhomogeneous and we quantify such an
inhomogeneuty by using the value of the standard temperature
deviation.

\section{The temperature standard deviation of the Bullet Cluster}

In this section, we calculate the temperature standard deviation
for the Bullet Cluster. This high temperature, merging galaxy
cluster is a good target for measuring the standard temperature
deviation because the contributions of high order corrections to
the SZ effect are strong and because the gas temperature
distribution is expected to be inhomogeneous along the line of
sight owing to the cluster merger effects. In the
following, we first describe the method to calculate the
temperature standard deviation along the line of sight that takes
into account SZ measurements at different frequencies and then
apply this method to the Bullet Cluster.

The relativistically correct SZ effect can be described in the
formalism based on the extension of the Kompaneets
equation which allows relativisitic effects to be included (e.g.,
Challinor \& Lasenby 1998; Itoh et al. 1998). Analytical forms for
the spectral changes due to the SZ effect, that are correct to
second order in the expansion parameter $\Theta=k_{\mathrm{b}}
T_{\mathrm{e}}/(m_{\mathrm{e}} c^2)$, can be written as

\begin{equation}
\frac{\Delta I(x)}{I_{\mathrm{0}}} \approx \sigma_{\mathrm{T}} \int
n_{\mathrm{e}} \Theta\times (g_0(x)+\Theta\times g_{1}(x)+\Theta^2\times g_{2}(x)) dl,
\label{form2}
\end{equation}
where $I_{\mathrm{0}}=2 (k_{\mathrm{b}} T_{\mathrm{cmb}})^3 /
(hc)^2$, $x=h\nu/k_{\mathrm{b}} T_{\mathrm{cmb}}$,
$T_{\mathrm{e}}$ is the electron temperature, $n_{\mathrm{gas}}$
is the number density of the ICM, $\sigma_{\mathrm{T}}$ is the
Thomson cross-section, $m_{\mathrm{e}}$ the electron mass, $c$ the
speed of light, $k_{\mathrm{b}}$ the Boltzmann constant, and $h$
the Planck constant. The spectral functions of $g_{0}(x)$,
$g_{1}(x)$, and $g_{2}(x)$ are taken from Eqs. (28) and (33) of
Challinor \& Lasenby (1998) and are given by

\begin{align}
&g_0(x)=\frac{x^4 \exp(x)}{(\exp(x)-1)^2} \left(x
\frac{\exp(x)+1}{\exp(x)-1}-4\right),
\label{g0}
\end{align}

\begin{align}
&g_{1}(x) = \frac{x^4 \exp(x)}{(\exp(x)-1)^2}\left(-10+\frac{47}{2}C(x)-\frac{42}{5}C^2(x)+\right.\nonumber\\
&\left.\frac{7}{10} C^3(x)+\frac{7x^2}{5\sinh^2(x/2)}\left(C(x)-3\right)\right),
\label{g1}
\end{align}

\begin{align}
&g_{2}(x)=\frac{x^4 \exp(x)}{(\exp(x)-1)^2}\left[-\frac{15}{2}+\frac{1023}{8} C(x)-\frac{868}{5} C^2(x)+
\right.\nonumber\\
&\left.\frac{329}{5} C^3(x)-\frac{44}{5} C^4(x)+\frac{11}{30} C^5(x)+\frac{x^2}{30\sinh^2(x/2)}\times
\right.\nonumber\\
&\left.\left(-2604+3948 C(x)-1452 C^2(x)+143 C^3(x)\right)\right.+\nonumber\\
&\left.\frac{x^4}{60\sinh^4(x/2)}\times\left(-528+187 C(x)\right)\right],
\label{g2}
\end{align}
where $C(x)=x\times\coth(x/2)$.

The spectral function, $g_{0}(x)$, corresponds to the spectral
function derived from the Kompaneets approximation (Zel'dovich \&
Sunyaev 1969). Both the first-order and second-order relativistic
effects make significant contributions to the spectral distortion
for $k_{\mathrm{b}}T_{\mathrm{e}}\gtrsim10$keV. We verified that,
for a plasma with the temperature of
$k_{\mathrm{b}}T_{\mathrm{e}}$ = 13.9 keV and optical depth $\tau
= 1.3\times$10$^{-2}$, as used by Zemcov et al. (2010), the
second-order corrections in $\tau$ to the spectral shape (see
Colafrancesco et al. 2003, for technical details) do not produce
appreciable variations in the SZ spectrum and, therefore, their
contribution will be neglected in our paper.

The CMB intensity change produced by the SZ effect from a galaxy
cluster with inhomogeneous density and temperature distributions
in the formalism based on an extension of the Kompaneets equation,
that is correct to second order in the expansion parameter
$\Theta$, is given by
\begin{align}
&\frac{\Delta I(x)}{I_{\mathrm{0}}} \approx \tau \left(
\frac{\langle k_{\mathrm{b}} T_{\mathrm{e}}\rangle}{m_{\mathrm{e}}
c^2} \times g_{0}(x)+\frac{\langle(k_{\mathrm{b}}T_{\mathrm{e}})^2\rangle}{m^2_{\mathrm{e}}
c^4}\times g_{1}(x)+\right.\nonumber\\
&\left.\frac{\langle(k_{\mathrm{b}}T_{\mathrm{e}})^3\rangle}{m^3_{\mathrm{e}}
c^6}\times g_{2}(x)\right),
\label{decomp}
\end{align}
where $\langle k_{\mathrm{b}} T_{\mathrm{e}}\rangle=\int
n_{\mathrm{e}}k_{\mathrm{b}}T_{\mathrm{e}} dl/\int n_{\mathrm{e}}
dl $ is the temperature averaged along the line-of-sight and
$\langle(k_{\mathrm{b}} T_{\mathrm{e}})^2\rangle=\int
n_{\mathrm{e}} (k_{\mathrm{b}}T_{\mathrm{e}})^2 dl/\int
n_{\mathrm{e}} dl $ is the squared temperature averaged along the
line-of-sight. Note that the optical depth of the electron plasma,
$\tau$, can be derived from X-ray observations.

To study the gas inhomogeneity along the line of sight, we
calculate the temperature standard deviation, $\sigma$, using the
method based on SZ intensity measurements (see Prokhorov et al.
2011).\\
The temperature standard deviation defined as
\begin{equation}
\sigma=\sqrt{\langle (k_{\mathrm{b}} T_{\mathrm{e}})^2\rangle-(\langle k_{\mathrm{b}} T_{\mathrm{e}}\rangle)^2}
\label{def}
\end{equation}
can be derived from the first two terms in Eq.
(\ref{decomp}) by using the SZ measurements at three or more
frequencies. The value of this quantity for a homogeneous gas
temperature distribution equals zero by definition. A finite,
non-zero value of $\sigma$ is therefore a measure of the
inhomogeneity of the gas temperature along the line of sight.

\begin{figure}
\centering
\includegraphics[angle=0, width=8.5cm]{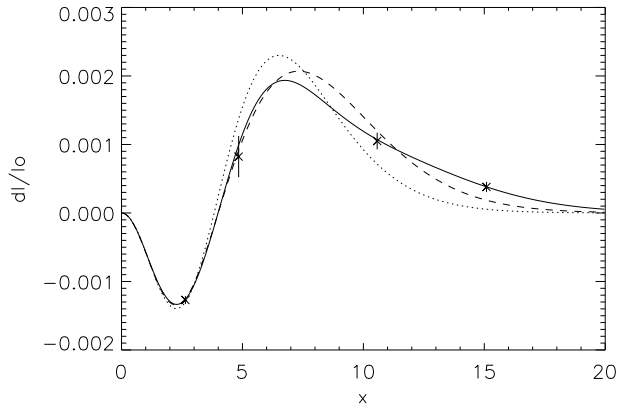}
\caption{The SZ spectrum of the Bullet Cluster fitted with the zero-order term (dotted),
with the zero plus first order terms (dashed), and with the zero, first, and second order terms (solid).}
\label{figure1}
\end{figure}

We derive the value of the temperature standard deviation by using
the available multifrequency SZ observations of the
Bullet Cluster: these have been obtained by ACBAR at
frequencies of 150 GHz and 275 GHz (Gomez et al. 2004) and by
Herschel-SPIRE at frequencies of 600 GHz and 857 GHz (Zemcov et
al. 2011). This set of SZ observations have been used by
Colafrancesco et al. (2011) in their studies of plasma models with
different thermal and non-thermal components.
The central SZ intensity values, $\Delta I/I_{0}$,
measured by ACBAR and Herschel-SPIRE are shown with errorbars in
Fig. \ref{figure1} and listed in Table \ref{tablesz} (in
units of MJy sr$^{-1}$). Note that the central SZ
intensity values have been derived by using the radial profiles
with an isothermal $\beta$ model (for a review, see Birkinshaw
1999) taking the core radius and $\beta$ parameter from Tucker et
al. (1998) for the ACBAR data (Gomez et al. 2004), and the density
profile given by Halverson et al. (2009) for the analysis of the
Hershel-SPIRE data (Zemcov et al. 2010). We checked the
consistence between the central SZ intensity values derived from
these models and found that the difference in the central values
is $\simeq$ 10\%. However, the gas temperature distribution should
be inhomogeneous along the line of sight and isothermal $\beta$
models are just an approximation used to fit the data. With that
caveat in mind, we will take the central SZ intensities from the
papers by Gomez et al. (2004) and Zemcov et al. (2010) as it has
been done in Colafrancesco et al. (2011).

\begin{table}
\begin{center}
\caption{The central values of the SZ effect at the
various frequencies as obtained by ACBAR and HERCSHEL-SPIRE}
\label{tablesz}
\begin{tabular}{@{}c c c c c}
\hline
Frequency (GHz) & $\Delta$ $I_{0}$ (MJy sr$^{-1}$) & Error (MJy sr$^{-1}$)\\
\hline
150 & -0.325 & 0.015\\
275 & 0.21 & 0.077\\
600 & 0.268  & 0.031 \\
857 & 0.097 & 0.019\\
\hline
\end{tabular}
\end{center}
\end{table}

The SZ intensity at each frequency is given by the sum of zero-th,
first, and second order contributions in $\Theta$ and can be
written as
\begin{equation}
\frac{\Delta I(x_{\mathrm{i}})}{I_{\mathrm{0}}} \approx A_{0}\times g_{0}(x_{\mathrm{i}})+A_{1}\times g_{1}(x_{\mathrm{i}})+A_{2}\times g_{2}(x_{\mathrm{i}})
\end{equation}
where the dimensionless frequencies $x_{1}$, $x_{2}$, $x_{3}$, and
$x_{4}$ equal to 2.64 ($\nu$=150 GHz), 4.84 ($\nu$=275 GHz), 10.57
($\nu$=600 GHz), and 15.09 ($\nu$=857 GHz), respectively. We use
the function
``MPFITEXPR''\footnote{\url{http://cow.physics.wisc.edu/~craigm/idl/idl.html}}
to find the best set of model parameters ($A_{0}$, $A_{1}$, and
$A_{2}$) which match the data performing Levenberg-Marquardt
least-squares fit. We also apply the same method to find the best
set of parameters in the models based on the Kompaneets
approximation and on the expression including the zero and first
order terms from Eq.(\ref{decomp}), respectively. We denote these
models as M1, M2, and M3, accordingly to the used number of terms
proportional to $\Theta$ for calculating the SZ effect
(i.e., the model M1 corresponds to the Kompaneets
approximation). The best-fit sets of parameters
($A_{0}$, $A_{1}$, and $A_{2}$) and 1-sigma errors on these
parameters ($\delta A_{0}$, $\delta A_{1}$, and $\delta A_{2}$)
derived from the SZ data are shown in Table 2 for the
models of M1, M2, and M3. The values of $\chi^2$ for these models
used to fit the observed SZ spectrum of the Bullet Cluster are
also shown in Table 2. Comparing the values of $\chi^2$
from Table 2 with those from the $\chi^2$ table, we
conclude that the model M3 agrees with the observed SZ data much
better than the models M1 and M2. The model M3 is consistent with
the SZ data and the $\chi^2$ value in this case is 0.3706 with 1
degree of freedom (d.o.f.). The SZ spectra derived from the models
M1, M2, and M3 are shown in Fig. \ref{figure1} by the
dotted, dashed, and solid lines, respectively. Comparing the SZ
spectra derived from the models with the observed SZ data, we find
that the high-frequency observations provided by
Herschel-SPIRE allow us to constain the contributions of the
first-order and second-order relativistic SZ corrections to the
thermal SZ effect. We calculate the covariance matrix for the
parameters, $A_{0}$, $A_{1}$, and $A_{2}$, in the model M3 using
the function of ``MPFITEXPR''. The calculated covariance matrix
is\\ ~\\ cov$_{\mathrm{ij}}$=$\begin{pmatrix} 2.44800E-10 &
1.42381E-11 & 1.67978E-13\\ 1.42381E-11 & 7.15626E-12 &
2.68991E-13\\ 1.67978E-13 & 2.68991E-13 & 3.38113E-14
\end{pmatrix}$\\ ~\\ where i, j = 0, 1, and 2. The square root of
the diagonal elements of the covariant matrix gives the 1-sigma
statistic errors on the parameters $A_{0}$, $A_{1}$, and $A_{2}$.

Using the definitions of the parameters of $A_{0}$ and $A_{1}$, we
rewrite Eq.(\ref{def}) as
\begin{equation}
\sigma=m_{\mathrm{e}} c^2 \sqrt{\frac{A_{1}}{\tau}-\left(\frac{A_{0}}{\tau}\right)^2}
\label{std}
\end{equation}

\begin{table}
\begin{center}
\caption{The values of parameters ($A_{0}$, $A_{1}$, and $A_{2}$), errors on these parameters, and $\chi^2$ for the
models M1, M2, and M3 used to fit the observed SZ spectrum.}
\label{table}
\begin{tabular}{@{}l c c c c}
\hline
Parameter & M1 & M2 & M3\\
\hline
$A_{0}$  & 3.338965E-4 & 3.66015E-4 & 3.68631E-4\\
$A_{1}$ & -- & 1.05000E-5 & 1.46551E-5\\
$A_{2}$ & -- & -- & 5.21968E-7\\
\hline
$\chi^2$  & 30.39981 & 8.42973 & 0.37056\\
d.o.f.  & 3 & 2 & 1\\
\hline
$\delta A_{0}$  & 1.45174E-5 & 1.56270E-5 & 1.56461E-5\\
$\delta A_{1}$ & -- & 2.23969E-6 & 2.6751E-6\\
$\delta A_{2}$ & -- & -- & 1.8488-7\\
\hline
\end{tabular}
\end{center}
\end{table}

We derive the value of the optical depth, $\tau$, using the X-ray
observations of the Bullet Cluster by ROSAT those have been
analyzed by Tucker et al. (1998). The surface brightness profile
of the primary cluster was obtained by extracting the net counts
in concentric annuli about the primary X-ray peak after excluding
the western quadrant (which contains the secondary peak) and all
point sources. Tucker et al. (1998) fitted the observed X-ray
surface brightness profile to the standard hydrostatic-isothermal
$\beta$ model and obtained the values of a core radius, a central
electron number density, and a beta parameter for the best fit. We
re-scale the values of a core radius, a central electron number
density using the present-day Hubble constant value, $H_{0}=74$ km
s$^{-1}$ Mpc$^{-1}$. Using the parameters of the beta model and
Monte-Carlo simulations, we find that the value of the optical
depth is $\tau=0.0138\pm0.0016$.

To calculate the temperature standard deviation along the line of
sight in the Bullet Cluster, we use the derived values of
parameters ($A_{0}$, $A_{1}$, and $A_{2}$), 1-sigma errors on
these parameters, covariance matrix (cov), and optical depth
($\tau=0.0138\pm0.0016$). The 1-sigma error on the standard
temperature deviation is given by
\begin{equation}
\delta \sigma=\sqrt{\sum_{i=0, 1}\left(\frac{\partial \sigma}{\partial A_{i}}\right)^2 \delta A^{2}_{i}+
2 \frac{\partial \sigma}{\partial A_{0}} \frac{\partial \sigma}{\partial A_{1}} \rm{cov}_{12}+
\left(\frac{\partial \sigma}{\partial \tau}\right)^2 \delta \tau^{2}}.
\label{root}
\end{equation}
We consider the value of the optical depth as derived
from X-ray observations, while the parameters ($A_{0}$, $A_{1}$,
and $A_{2}$) are derived from SZ observations. Thus, the optical
depth is an uncorrelated variable with other parameters in
Eq.(\ref{root}).

Using Eqs.(\ref{std}) and (\ref{root}), we find that the value of
temperature standard deviation in the Bullet Cluster equals to
$\sigma=9.5\pm2.6$ keV. This value of standard temperature
deviation is high although somewhat lower than the average plasma
temperature, $k_{\mathrm{b}} T_{\mathrm{e}}\approx 14.5$ keV,
derived from the Chandra X-ray observations of the Bullet Cluster
(see, e.g., Million \& Allen 2009).

Since the derived value of temperature standard deviation is high
and the contribution of higher order terms in the expansion
parameter $\Theta$ is significant at high frequencies for
temperatures $\simgt$ 15 keV, we recalculate the values of the
temperature standard deviation and 1-sigma error on this quantity
considering the spectral changes due to the SZ effect correct to
third order in $\Theta$. We estimate the uncertainty of the
technique comparing the SZ intensity value derived in the extended
Kompaneets formalism with that is derived from the Wright
formalism (1979) and take them into account as systematic errors
in the analysis. In this analysis, we consider electron
temperatures in the range 5 keV -- 30 keV that agree with
that was found in the recent hydrodynamic simulations of the
Bullet Cluster (Akahori \& Yoshikawa 2011). We find that the
values of the temperature standard deviation and 1-sigma error,
$\sigma=10.6\pm3.8$ keV, are consistent with those derived above.

We also calculated the temperature standard deviation
using the optical depth derived by using the radial density
profiles given by Halverson et al. (2009) and Ota \& Mitsuda
(2004). We obtained the values of the temperature standard deviation and of the  
associated errors that are within 10\% and 30\%, respectively, of those  
derived above. Therefore, the temperature standard
deviation is high compared to the value of the gas temperature
also assuming these models for the radial density distribution.

Based on our results, we conclude that the
plasma temperature distribution is very inhomogeneous along the
line of sight in the Bullet Cluster. This result confirms the
presence of additional electron components previously revealed by
Million \& Allen (2009) by means of X-ray observations and by
Colafrancesco et al. (2011) by means of SZ observations. Note that
the method of Prokhorov et al. (2011), that we use in this paper,
can be also applied to other merging massive galaxy clusters as
soon as high-frequency SZ observations of other clusters become
available.

\section{Conclusions}

We have derived the value of temperature standard deviation,
$\sigma$, along the line of sight in the Bullet Cluster using SZ
effect observations at frequencies of 150, 275, 600, and 857 GHz.
The measured value, $\sigma=9.5\pm2.6$ keV, is obtained from the
analysis taking into account the three first terms in the
expansion parameter $\Theta=k_{\rm{b}} T_{\rm{e}}/m_{\rm}c^2$ of
the SZ effect (see Eq. \ref{form2}), and the analysis
including the four first terms in $\Theta$ and taking into account
the uncertainty of the technique gives a consistent value,
$\sigma=10.6\pm3.8$ keV. The derived value of $\sigma$ is of the
order of the average plasma temperature, $k_{\mathrm{b}}
T_{\mathrm{e}}=14.5$ keV, in this cluster. Such a high value of
$\sigma$ suggests hence that the gas temperature
distribution is strongly inhomogeneous along the line of sight.

The high plasma temperature inhomogeneity along the line of sight
in the Bullet Cluster is likely caused by its past merger
activity. The presence of the strong shock wave and ``bullet''
cold substructure in this galaxy cluster is strong evidence in
favor of a recent major merger of subclusters. Our result
shows that the value of temperature standard deviation along the
line of sight can be quite high in massive clusters
undergoing strong merging and, therefore, that the plasma in such
clusters should be strongly disturbed. This agrees with the fact
that a single temperature model does not provide a good fit to
X-ray observations (see Million \& Allen 2009) and to SZ effect
measurements (see Colafrancesco et al. 2011). By comparing
the derived temperature standard deviation for the different
models of a radial density distribution, we verified that our
results are insensitive to the choice of the radial density
profile. However, note that our results can be subject to
additional possible systematic errors that come from:
i) the uncertainties in the flux scales between the different
frequency bands;
ii) incomplete knowledge of the radial gas density distribution;
iii) the subtraction of the bright sub-mm sources that can cause
confusion in the ACBAR as well as the SPIRE data (see Johansson et
al. 2010).
The first two of these issues will be considered in a following
paper (Prokhorov et al. 2012) where a joint analysis of the X-ray
and SZ data for the Bullet Cluster will be presented.

The technique we present here does not require any
specific assumptions about temperature profile. Therefore, the
advantage of using the temperature standard deviation is that this
quantity provides a model-independent information of the plasma temperature 
inhomogeneity along the line of sight.

There are several astrophysical and cosmological consequences of
having large values of temperature standard deviation, $\sigma$,
in galaxy clusters:\\
i) measuring $\sigma$ in different clusters of galaxies will tell
us the relevance of merging and shocks in the evolution of these
systems, especially those undergoing a collision along the
line-of-sight;\\
ii) the specific value of $\sigma$ will allow us to
constrain the possible non-gravitational effects in clusters and
their evolution with the mass and redshift of the system. In
addition it will tell us how much of the energy density of the
cluster is due to non-gravitational effects with respect
to pure gravitational effects;\\
iii) the value of $\sigma$ will tell us how confidently we can use
galaxy clusters as cosmological probes (e.g., those are based on
using of a mass-temperature relation) assuming simple models for
the ICM, such a single temperature model.\\
We will discuss these and other consequences in more details
elsewhere.

The first measurement of temperature standard deviation along the
line-of-sight, that we have presented in this paper, demonstrates
that existing microwave and mm. instruments provide us with
a unique opportunity to derive this quantity in massive,
merging clusters. Future multi-frequency SZ effect observations
including high frequency observations of other merging galaxy
clusters will allow us to study the plasma temperature
distribution along the line of sight for many clusters of galaxies
and will provide us with a better understanding of gas dynamical
processes occurring in the ICM of merging clusters.

This is the first attempt in which we have been successful to
measure $\sigma$, and we plan to provide a systematic analysis of
the temperature standard deviation in all clusters with sensitive
enough multi-frequency SZE observations. We believe that this new
tool will provide a promising method to analyze the complex
structure of cluster atmospheres.

\section{Acknowledgements}

We are grateful to Shigehiro Nagataki, Dmitry Malyshev, and Andrey
Vladimirov for discussions and we thank the Referee for
valuable suggestions.

\label{lastpage}

\end{document}